\documentclass[11pt]{article}

\usepackage[T1]{fontenc}
\usepackage{amsmath}
\usepackage{graphicx}
\usepackage{indentfirst}
\usepackage{amssymb}
\usepackage{cite}
\usepackage{color}
\usepackage{subfigure}
\usepackage{varwidth}
\usepackage{subfigure}
\usepackage[colorlinks=true, linkcolor=red, citecolor=blue, urlcolor=magenta]{hyperref}
\usepackage{xcolor}
\usepackage{amsmath}
\usepackage{bm}
\usepackage[normalem]{ulem}

\begin{document}
	
\title{Distinguishing Black Holes and Neutron Stars via Optical Images Illuminated by Thick Accretion Disks}

\date{}
\maketitle

\begin{center}
\author{Chen-Yu Yang,}$^{a}$\footnote{E-mail:  chenyu\_yang2024@163.com}
\author{Xiao-Xiong Zeng}$^{b}$\footnote{E-mail: xxzengphysics@163.com (Corresponding author)}
\\

\vskip 0.25in
$^{a}$\it{Department of Mechanics, Chongqing Jiaotong University, Chongqing 400000, People's Republic of China}\\
$^{b}$\it{College of Physics and Optoelectronic Engineering, Chongqing Normal University, Chongqing 401331, People's Republic of China}\\
\end{center}

\vskip 0.6in
{\abstract{
This paper investigates the optical images of neutron stars within the framework of the radiatively inefficient accretion flow model, taking into account a polytropic equation of state. After obtaining the numerical solutions of the neutron star, we solved numerically the geodesic equations together with the radiative transfer equation. We mainly examine the effects of the polytropic index $N$ and the observer inclination angle $\theta_o$ on the image morphology. The obtained images are also compared with the shadow of a Schwarzschild black hole. It is shown  that, under the assumption that photon trajectories are terminated at the neutron star surface, the image exhibits a bright higher order structure surrounding an inner dark region. As $N$ increases, the size of the higher order image gradually expands. As $\theta_o$ increases, the obscuration of the neutron star silhouette by radiation originating outside the equatorial plane becomes more pronounced. Compared with the black hole shadow obtained under the same parameter configuration, the neutron star exhibits a larger higher order image and a more extended obscured inner dark region, whereas the higher order image of the black hole is more readily distinguishable. These results indicate significant differences in the optical appearance of neutron stars and black holes, and thus provide a theoretical basis for distinguishing between them through high resolution imaging.
}}

\thispagestyle{empty}
\newpage
\setcounter{page}{1}


\section{Introduction}
Neutron stars are formed through the gravitational collapse of the cores of massive stars with masses exceeding $8 M_\odot$ at the end of their lives. This process usually triggers a Type II supernova explosion. Newly born neutron stars are rich in leptons, mainly electrons $e^{-}$ and electron neutrinos $\nu_e$. Although the detailed mechanism of Type II supernova explosions is not yet fully understood~\cite{burrows2000supernova}, neutrinos are widely believed to play a crucial role in this process. A notable feature is that, during the collapse, neutrinos are temporarily trapped inside the star. When the density in the stellar interior reaches the nuclear saturation density $n_0$, the core collapse is halted, and a shock wave is formed at the outer edge of the core. After propagating outward to a distance of about $100$--$200\,\mathrm{km}$, the shock wave stalls because it loses energy through neutrino emission as it moves outward. With the assistance of rotation, convection, and magnetic fields, neutrinos emitted from the core may revive the stalled shock wave, causing it to accelerate outward again within a few seconds and eject the mantle of the massive star~\cite{burrows1986birth,bionta1987observation,burrows1987neutrinos}.

In some cases, the proto-neutron star may fail to remain stable during its early evolution and instead undergo further collapse into a black hole. The proto-neutron star accretes fallback material passing through the shock wave. This accretion process ceases once the shock wave resumes its outward propagation. Before that happens, however, the stellar mass may already have exceeded the maximum allowed mass. In such a case, the proto-neutron star will collapse~\cite{burrows1988supernova}. Even if this does not occur, there exists a second possible mechanism for black hole formation~\cite{prakash2001neutrino}. Compared with a cold star, the proto-neutron star can sustain a higher maximum mass because of its additional lepton content and thermal energy. Consequently, after accretion has ended, the mass of the proto-neutron star may be below its own maximum mass while still exceeding the maximum mass allowed for a cold star. In this case, the proto-neutron star will collapse into a black hole on a diffusion timescale of $10$--$20\,\mathrm{s}$.

An isolated neutron star will eventually exhaust its thermal and magnetic energy and gradually become dim. However, when a neutron star is in a binary system, its evolutionary process becomes much more complex. For example, the neutron star may accrete a large amount of material from its companion, and the gravitational and thermonuclear energy released in this process can make it a bright source of X-ray radiation~\cite{potekhin2010physics,lipunov1992astrophysics}. In such a case, the accreted material forms an accretion disk around the neutron star. The accretion disk itself also emits X-rays, and the luminosity of the neutron star evolves over time because of disk precession or variations in the accretion rate. Overall, the presence of an accretion disk can significantly alter the optical image of the neutron star.

When a compact object is illuminated by its own accretion disk, strong gravitational lensing gives rise to a bright ring-like structure together with a dim central region~\cite{luminet1979image,falcke1999viewing,wambsganss1998gravitational,yang2025observational}. This phenomenon is consistent with the radiation image of the overheated plasma surrounding the supermassive compact object at the center of M87 obtained by the Event Horizon Telescope (EHT)~\cite{akiyama2019first1,akiyama2019first2,akiyama2019first3,akiyama2019first4,akiyama2019first5,akiyama2019first6}. On the other hand, different compact objects may exhibit significant differences in their photon sphere structures. For example, some may possess multiple photon rings, whereas others may have no photon ring at all~\cite{wielgus2020reflection,tsukamoto2021gravitational,olmo2022new,guerrero2022light,tsukamoto2022retrolensing}. These structural differences provide a theoretical basis for distinguishing different compact objects through their optical images.

At present, substantial progress has been made in the study of the optical images of compact objects, especially black hole shadows\cite{zeng2022effects,zeng2023optical,zeng2022qed,chen2025polarization,hou2024new,zhang2024imaging,zhao2026probing,wang2026geodesics,fan2025magnetic}. Existing works usually treat the accretion disk as the primary radiation source and have developed a variety of accretion models~\cite{narayan2019shadow,zeng2020shadows,zeng2022shadows,li2021shadows,hashimoto2020imaging,aslam2024holographic,yang2026observational,zeng2025schwarzschild,li2026observational}. For the sake of numerical simplicity, geometrically thin and optically thin accretion disk models have been widely adopted in studies of black hole shadow imaging\cite{hou2022image}. Such models can capture key features such as the inner shadow and the critical curve~\cite{zeng2025holographic}. However, EHT and related observations indicate that, in strong gravity environments, the accretion flow around supermassive black holes may evolve into a geometrically thick and optically thin structure because of the suppression of vertical cooling and the compression of matter~\cite{akiyama2019first1,akiyama2019first2,akiyama2019first3,ho1999spectral}. This implies that the accreting material is not confined to the equatorial plane, and it is therefore insufficient to consider only the instantaneous emission and absorption of photons within that plane. Under such circumstances, it is necessary to incorporate the effects of key physical factors such as the electron number density, electron temperature, and magnetic-field structure. In studies of geometrically thick accretion disks, most previous works have adopted the phenomenological radiatively inefficient accretion flow (RIAF) model. After vertical averaging, the electron number density and electron temperature are usually assumed to follow approximate power-law distributions with the radial coordinate~\cite{yuan2003nonthermal,broderick2005frequency,broderick2009imaging,wielgus2025semi,jiang2024shadows,yang2026shadow}. The RIAF model has successfully reproduced the overall image features of M87*~\cite{akiyama2019first5} and is in close agreement with general relativistic magnetohydrodynamic (GRMHD) simulations, providing a solid framework to discuss the properties of thick, low radiation efficiency disks.

RIAF, standard and normal evolution (SANE), and magnetically arrested disk (MAD) are not mutually exclusive at the conceptual model level. In this work, we adopt the RIAF framework to describe the thermodynamic and radiative properties of low accretion rate flows. These flows are hot, optically thin, geometrically thick, and radiatively inefficient. By contrast, SANE and MAD represent two limiting GRMHD configurations of magnetized thick RIAF flows, corresponding to weakly or moderately magnetized states (SANE) and strongly magnetized states (MAD), respectively. Specifically, SANE corresponds to flows in which the magnetic flux remains below saturation and the magnetic field primarily mediates angular momentum transport, whereas MAD corresponds to flows in which the magnetic flux approaches saturation near the compact object, strongly modifying the flow dynamics and often producing more powerful jets and outflows \cite{narayan1994advection,yuan2014hot,narayan2012grmhd,narayan2003magnetically,tchekhovskoy2011efficient,skadowski2013energy}. Our aim is to isolate the effect of finite disc thickness on optical images, rather than modeling magnetic flux accumulation, jet launching, or strong GRMHD variability. For this purpose, a parameterized RIAF provides the most appropriate baseline model because it retains the essential thermodynamic structure expected for low accretion rate flows while avoiding the additional assumption that the flow must already be in either the SANE or MAD magnetic flux limit. Moreover, the RIAF framework captures the key features of optical images, including higher order images and obscuration effects, while allowing calculations to be performed with relative simplicity and ensuring numerical stability. This choice is also supported by current studies of weakly magnetized neutron star systems. Observations of Cen X‑4 indicate that the accretion flow can be radiatively inefficient and does not require strong magnetic regulation \cite{d2015radiative}. Furthermore, studies of advection-dominated accretion flows (ADAFs) in weakly magnetized neutron stars show that even in the presence of a hard surface, the inner hot flow remains radiatively inefficient, supporting the adoption of a RIAF approximation \cite{qiao2020systematic,qiao2021radiative}.

Although studies of black hole shadows have become extensive, investigations of the optical images of neutron stars remain relatively limited. In our previous work, we studied the optical images of neutron stars illuminated by a thin accretion disk~\cite{yang2026distinguishing}. The results showed that, under the assumption that photon trajectories are terminated at the neutron star surface, the maximum intensity in the neutron star image appears at the stellar surface, whereas in the black hole image it appears near the photon ring. In addition, for the same mass, the region of reduced intensity in the interior is more extended for the neutron star than for the black hole. These results indicate that neutron stars and black holes exhibit clear differences in their optical imaging features. Geometrically thick and optically thin accretion structures of the kind described above may also arise in neutron star systems. Motivated by this possibility, this paper adopts neutron star models with a polytropic equation of state~\cite{flanagan2008constraining,lattimer2007neutron,hinderer2008tidal,read2009constraints} and systematically investigates their optical images within the RIAF framework, in comparison with black hole shadows. This study provides a theoretical basis for distinguishing neutron stars from black holes.

The structure of this paper is as follows. Sec.~\ref{sec2} briefly introduces the construction of the equilibrium equations for the neutron star interior and the ray-tracing method. Sec.~\ref{sec3} presents the structural properties of geometrically thick accretion flows and the electron radiation mechanism. Sec.~\ref{sec4} shows the numerical results. Sec.~\ref{sec5} summarizes the paper and provides a discussion. Unless otherwise specified, geometrized units are used throughout this paper, with $c=G=1$, where $c$ is the speed of light in vacuum and $G$ is the gravitational constant.

\section{Equilibrium Structure and Ray-Tracing Framework}\label{sec2}
The spacetime of a static, spherically symmetric star can be described by the following metric~\cite{misner1973gravitation}
\begin{equation}
	ds^2 = g_{\mu\nu} dx^\mu dx^\nu
	= - e^{A(r)} dt^2 + e^{B(r)} dr^2 + r^2 d\Omega^2,
	\label{eq:ssm}
\end{equation}
where $d\Omega^2 = d\theta^2 + \sin^2\theta \, d\phi^2$ is the line element of the unit two sphere. The matter field inside the star can be approximately treated as an ideal fluid, whose stress energy tensor is
\begin{equation}
	T_{\mu\nu} = (\rho + p) u_{\mu} u_{\nu} + p g_{\mu\nu},
\end{equation}
where $u^\mu$ is the fluid four velocity, and $\rho$ and $p$ denote the energy density and pressure of the fluid, respectively. From the Einstein field equations, one can derive the Tolman Oppenheimer Volkoff (TOV) equation
\begin{equation}
	p' = -(\rho + p)\frac{m(r) + 4\pi p r^{3}}{r\left[r - 2m(r)\right]}.
	\label{eq:tov}
\end{equation}
In addition,
\begin{align}
	m(r) &= 4\pi \int_{0}^{r} \rho(x) x^{2} \, dx, \\
	m'(r) &= 4\pi \rho(r) r^{2}, \\
	A'(r) &= \frac{2\left[m(r) + 4\pi p r^{3}\right]}{r\left[r - 2m(r)\right]}, \\
	e^{B(r)} &= \left[1-\frac{2m(r)}{r}\right]^{-1}.
	\label{eq:tov2}
\end{align}
Here the prime denotes differentiation with respect to the radial coordinate $r$. For compact objects such as neutron stars, the equation of state can be written in the general form $f(p,\rho)=0$~\cite{liang2023differential}. In this paper, we adopt a polytropic equation of state~\cite{hinderer2009erratum,read2009constraints}
\begin{equation}
	p - k \rho^\gamma = 0,
\end{equation}
where $k$ is a constant and $\gamma$ is the adiabatic index, given by
\begin{equation}
	\gamma = 1 + \frac{1}{N},
\end{equation}
with $N$ being the polytropic index. By solving Eqs.~(\ref{eq:tov})--(\ref{eq:tov2}), one can obtain the neutron star radius $R_*$ and the total mass $M_* = m(R_*)$. At the same time, the numerical solutions for the interior metric functions can also be determined. In the exterior region of the star, the Schwarzschild metric is adopted. The matching condition at the stellar surface $r=R_*$ is
\begin{equation}
	e^{A(R_*)} = 1 - \frac{2M_*}{R_*}.
\end{equation}
It should be noted that directly using the numerically obtained metric in the subsequent imaging calculations is not particularly convenient. Therefore, a fitted metric is adopted in the following analysis. The detailed procedure can be found in Ref.~\cite{rosa2022shadows} and in our previous works~\cite{yang2026distinguishing,he2025observation}.

We next derive the equations of motion for photons propagating in the vicinity of a neutron star. For convenience, metric~(\ref{eq:ssm}) is rewritten as
\begin{equation}
	ds^2 = -\tilde{A}(r) dt^2 + \tilde{B}(r)^{-1} dr^2 + r^2 d\Omega^2,
\end{equation}
where $\tilde{A}(r)=e^{A(r)}$ and $\tilde{B}(r)=e^{-B(r)}$. The photon trajectories satisfy the Euler--Lagrange equations
\begin{equation}
	\frac{d}{d\tau}\left(\frac{\partial L}{\partial \dot{x}^\mu}\right) = \frac{\partial L}{\partial x^\mu},
\end{equation}
where $\dot{x}^{\mu}$ denotes the tangent vector of the photon trajectory, the dot represents differentiation with respect to the affine parameter $\tau$, and $L$ is the Lagrangian. Taking $x^\mu=\{t,r,\theta,\phi\}$, the Lagrangian can be written as
\begin{align}
	L &= \frac{1}{2} g_{\mu\nu} \dot{x}^{\mu} \dot{x}^{\nu} \notag \\
	&= \frac{1}{2} \left[
	-\tilde{A}(r)\dot{t}^{2}
	+ \tilde{B}(r)^{-1}\dot{r}^{2}
	+ r^{2}\dot{\theta}^{2}
	+ r^{2}\sin^{2}\theta\,\dot{\phi}^{2}
	\right] \notag \\
	&= 0 .
	\label{eq:lag}
\end{align}
Since this static and spherically symmetric metric does not depend explicitly on the time coordinate $t$ or the azimuthal angle $\phi$, the spacetime admits two conserved quantities. Restricting null geodesics to the equatorial plane $\theta=\pi/2$, one obtains
\begin{equation}
	\tilde{E} = -\frac{\partial L}{\partial \dot{t}}
	= \tilde{A}(r)\dot{t}, \qquad
	\tilde{L} = \frac{\partial L}{\partial \dot{\phi}}
	= r^2 \dot{\phi},
	\label{eq:el}
\end{equation}
which correspond to the conserved energy and angular momentum, respectively. Defining the impact parameter as $\tilde{I} \equiv |\tilde{L}|/\tilde{E}$ and setting $|\tilde{L}|=1$, one obtains from Eqs.~(\ref{eq:lag}) and~(\ref{eq:el}) the following components of the photon four velocity
\begin{align}
	\dot{t} &= \frac{1}{\tilde{I}\tilde{A}(r)}, \label{eq:geo1} \\
	\dot{r} &= \sqrt{
		\frac{1}{\tilde{I}^2}\frac{\tilde{B}(r)}{\tilde{A}(r)}
		- \frac{\tilde{B}(r)}{r^2}
	}, \\
	\dot{\theta} &= 0, \\
	\dot{\phi} &= \frac{1}{r^2}.
	\label{eq:geo4}
\end{align}

The geodesic Eqs.~(\ref{eq:geo1})--(\ref{eq:geo4}) accurately describe photon propagation in the vicinity of a neutron star. To specify the initial conditions of the photon trajectories, it is also necessary to introduce an observer. In this paper, we adopt a zero angular momentum observer (ZAMO). Suppose that the observer is located at the coordinate position $(t_o, r_o, \theta_o, \phi_o)$. Then, in the neighborhood of the observer, one can construct a local orthonormal tetrad as
\begin{equation}
	e_{(a)}^{\ \beta} = \operatorname{diag}
	\left(
	\frac{1}{\sqrt{-g_{tt}}},
	-\frac{1}{\sqrt{g_{rr}}},
	\frac{1}{\sqrt{g_{\theta\theta}}},
	-\frac{1}{\sqrt{g_{\phi\phi}}}
	\right).
\end{equation}
In this local reference frame, the photon four-momentum is given by
\begin{equation}
	p_{(a)} = e_{(a)}^{\ \beta} p_{\beta},
\end{equation}
where $p_\beta$ is the four momentum in the Boyer--Lindquist coordinate system. On this basis, the celestial coordinates $(X,Y)$ of the observer are introduced, and their relations to the components of the four momentum are
\begin{equation}
	\cos X = \frac{p^{(1)}}{p^{(0)}}, 
	\quad
	\tan Y = \frac{p^{(3)}}{p^{(2)}}.
\end{equation}
Cartesian coordinates $(x,y)$ are then introduced on the observer's screen~\cite{hu2021qed}, and they satisfy
\begin{equation}
	x = -2 \tan\frac{X}{2}\sin Y, 
	\quad 
	y = -2 \tan\frac{X}{2}\cos Y.
	\label{eq:co1}
\end{equation}
For numerical imaging, the image plane is divided into $n \times n$ pixels, each labeled by $(\tilde{X},\tilde{Y})$~\cite{yang2025shadow}. The relation between the Cartesian coordinates $(x,y)$ and the pixel coordinates is
\begin{equation}
	x = \ell \left(\tilde{X}-\frac{n+1}{2}\right), 
	\quad 
	y = \ell \left(\tilde{Y}-\frac{n+1}{2}\right),
	\label{eq:co2}
\end{equation}
where $\ell$ is the side length of a single pixel. Combining Eqs.~(\ref{eq:co1}) and~(\ref{eq:co2}), one obtains the mapping between the pixel coordinates and the celestial coordinates
\begin{align}
	\tan\frac{X}{2} 
	&= \frac{1}{n}\tan\left(\frac{\gamma_{\mathrm{fov}}}{2}\right)
	\left[\left(\tilde{X}-\frac{n+1}{2}\right)^2+\left(\tilde{Y}-\frac{n+1}{2}\right)^2\right]^{\frac{1}{2}}, 
	\label{eq:c1} \\
	\tan Y 
	&= \frac{2\tilde{Y}-(n+1)}{2\tilde{X}-(n+1)},
	\label{eq:c2}
\end{align}
where $\gamma_{\mathrm{fov}}$ is the field of view~\cite{he2026shadow}.

It should be emphasized that, in the actual ray-tracing procedure, the neutron star radius $R_*$ is taken as one of the termination conditions for the integration. Once the radial coordinate of a photon reaches the stellar surface $r=R_*$, the numerical integration is stopped. This treatment is based on the fact that the interior of a neutron star consists of extremely dense nuclear matter and is effectively optically thick to electromagnetic radiation~\cite{lattimer2001neutron,lattimer2004physics}. Therefore, this paper considers only photon propagation in the exterior spacetime of the star, with the main focus on the influence of spacetime curvature on the geodesic structure rather than on the full radiative transfer process. Physically, the stellar radius $R_*$ is equivalent to an optically thick boundary. In the numerical implementation, this treatment avoids integration results that would lack physical significance if photons were allowed to continue propagating into regions of extremely strong spacetime curvature, thereby improving the stability of the imaging calculations. Under the thin accretion disk model, this boundary condition gives rise to a dark region in the central area. When a geometrically thick accretion disk is considered, however, the radiation distribution away from the equatorial plane may significantly alter the optical structure of the neutron star.

We further clarify the physical meaning of the integral truncation condition. In general, the stellar surface at $r=R_*$ does not correspond to a turning point for radial photon motion, and photon trajectories are not naturally reversed by the geodesic effective potential at this location. In the present calculation, photons reaching this boundary are removed from the external radiative transfer computation. They are not traced further for surface processes such as reflection, transmission, or delayed re-emission, and therefore no longer contribute to the observed intensity on the image plane. In this sense, an effective absorbing boundary condition is adopted. Numerically, this is equivalent to assuming that the neutron star surface fully absorbs incoming photons. This treatment can also be regarded as an idealized approximation of the physics at the neutron star surface. The actual surface or near-surface region may involve more complex processes. If such processes were taken into account, part of the energy reaching the surface might eventually escape. However, internal propagation, matter interactions, and the strong gravitational field near the surface could introduce significant time-delay effects. These effects would reduce the instantaneous contribution of such photons to the observed image and produce a central decrease in intensity similar to that found in this work. A full modeling of these surface processes is beyond the scope of this work. Therefore, the central dark region discussed in this paper should be understood as a feature arising from the integration truncation under the adopted effective boundary condition.

Similar issues have also been discussed in studies of horizonless compact objects with physical surfaces. Early work investigating the event horizon of M87 indicated that, if the central compact object were not a black hole but instead had a physical surface, this surface could produce considerable thermal near-infrared and optical emission~\cite{broderick2015event}. Related EHT studies of Sgr~A* also considered compact object alternatives with surfaces that thermally re-emit incident radiation or partially reflect it. They further concluded that, for Sgr~A*, a thermal surface can be ruled out, while a fully reflective surface is unlikely~\cite{event2022first}. However, Refs.~\cite{carballo2022constraints,carballo2023constraints} pointed out that observational constraints based on prompt thermal re-emission from a surface rely on strong physical assumptions. In particular, they require that the energy exchange between the central object and the accretion flow reaches equilibrium on a sufficiently short timescale and that the accretion energy is re-emitted in the relevant observational bands. For a general compact object with a physical surface, the interaction between the incident energy and the central object may involve different channels, including temporary absorption, instantaneous re-emission, reflection, and transmission. The existence of a physical surface does not necessarily produce a significant brightness contribution in the relevant observational band or timescale. Therefore, the boundary condition adopted here can be understood as an effective description of the external radiative transfer in steady state, rather than a simulation of the full microphysical processes at the neutron star surface.

\section{Geometrically Thick Accretion Flow Model}\label{sec3}
In this paper, we consider a geometrically thick and optically thin accretion flow model around a neutron star, namely the phenomenological RIAF model. Unlike the standard thin disk model, a geometrically thick accretion flow requires the explicit specification of physical quantities such as the particle number density, electron temperature, and magnetic field configuration. In principle, these quantities can be obtained self-consistently by solving the GRMHD equations. However, because of the high complexity of the GRMHD equations, this section adopts a numerical approach with appropriate simplifications.

\subsection{Flow Structure and Dynamics}
In cylindrical coordinates, the radius is defined as $R = r \sin\theta$ and the height as $z = r \cos\theta$, with $z=0$ on the equatorial plane $\theta = \pi/2$. The electron number density and electron temperature distributions are given by~\cite{quataert2003radiatively,igumenshchev2003three}
\begin{equation}
	N_e = N_*\left(\frac{R_*}{r}\right)^2 
	\exp\left(-\frac{z^2}{2 R^2}\right), 
	\quad
	T_e = T_*\left(\frac{R_*}{r}\right),
\end{equation}
where $N_*$ and $T_*$ denote the electron number density and electron temperature at the neutron star surface $r = R_*$, respectively. Here, the neutron star radius $R_*$ is taken as the inner boundary scale, replacing the event horizon radius $r_h$ used in black hole models. The magnetic-field strength is defined as
\begin{equation}
	b = \sqrt{\sigma \rho},
\end{equation}
where $\sigma \sim 0.1$ is the cold magnetization parameter~\cite{pu2016effects,johnson2015resolved}, and $\rho = N_e m_p c^2$ denotes the rest-mass energy density of the fluid, with $m_p$ and $c$ being the proton mass and the speed of light in vacuum, respectively.

In the accretion flow model, this paper considers the fluid to undergo a purely infalling motion. Assuming that the fluid is at rest at infinity, namely $u_t = -1$, its four velocity can be written as~\cite{takahashi2011constraining,vincent2022images}
\begin{equation}
	u^\mu = \left(-g^{tt}, -\left[-(1+g^{tt}) g^{rr}\right]^{\frac{1}{2}}, 0, 0 \right).
\end{equation}
These expressions will be used in the subsequent radiative transfer calculation. We also tested orbiting and combined motion, and the comparison among the three accretion flow motion modes is presented in Appendix~\ref{appendix1}.

\subsection{Radiative Transfer and Synchrotron Emission}
In the unpolarized case, the radiative transfer equation for the corresponding invariant quantities is
\begin{equation}
	\frac{d}{d\tau}\hat{I}=\hat{J}-\hat{\alpha}\hat{I}.\label{eq:rte}
\end{equation}
In terms of the physical quantities, it can be written as
\begin{equation}
	\frac{d}{d\tau}\frac{I_{\nu}}{\nu^{3}}=\frac{j_{\nu}}{\nu^{2}}-(\nu\alpha_{\nu})\frac{I_{\nu}}{\nu^{3}},
\end{equation}
where $I_{\nu}$, $j_{\nu}$, and $\alpha_{\nu}$ denote the specific intensity, emissivity, and absorptivity at photon frequency $\nu$ in the local reference frame, respectively. The solution to Eq.~(\ref{eq:rte}) in geometrized units is
\begin{equation}
	\hat{I}(\tau)=\hat{I}(\tau_0)+\int_{\tau_0}^\tau d\tau^{\prime}\hat{J}(\tau^{\prime})\exp\left(-\int_{\tau^{\prime}}^\tau d\tau^{\prime\prime}\hat{\alpha}(\tau^{\prime\prime})\right).
\end{equation}
In our ray-tracing calculations, geometrized units with $G = c = 1$ are used, and the reference photon is assigned a frequency $\nu = 1$ at infinity. However, the radiation quantities $I_\nu$, $j_\nu$, and $\alpha_\nu$ are defined in CGS units, so the two sets of quantities are in different unit systems. To consistently express all quantities in CGS units, we introduce a scaling factor $K = r_g / \nu_0$, where $r_g = GM_*/c^2$ is the unit length, and $\nu_0$ is the photon frequency measured at infinity, both defined in CGS units. Multiplying Eq.~(\ref{eq:rte}) by $K$ scales the affine parameter $\tau$ such that $d\tau_{\rm phys} = K\, d\tau_{\rm geom}$ corresponds to a physical distance in CGS units. This ensures that the subsequent integration along the photon path produces intensities with the correct physical units. Eq.~(\ref{eq:rte}) then becomes
\begin{equation}
	\frac1{K}\frac{ d}{ d\tau}\hat{I}=\hat{J}-\hat{\alpha}\hat{I},
\end{equation}
with the solution
\begin{equation}
	I_{\nu}=\hat{g}^{3}I_{\nu_{0}}+r_{g}\int_{\tau_{0}}^{\tau} d\tau^{\prime}\hat{g}^{2}j_{\nu}(\tau^{\prime})\exp\left(-r_{g}\int_{\tau^{\prime}}^{\tau} d\tau^{\prime\prime}\alpha_{\nu}(\tau^{\prime\prime})/\hat{g}\right),\label{eq:is}
\end{equation}
where $\hat{g}=\nu_{0}/\nu$ is the redshift factor. Let $k_\mu$ be the photon four-momentum. Then
\begin{equation}
	\hat{g}=\frac{k_t}{k_\mu u^\mu}=\frac{-1}{k_\mu u^\mu}.
\end{equation}
The local magnetic field $b^\mu$ satisfies the orthogonality condition $b_{\mu} u^\mu = 0$. It is therefore clear that the key to the numerical calculation of the intensity is the accurate determination of the emissivity $j_\nu$ and the absorptivity $\alpha_\nu$.

For the remainder of this subsection, the CGS unit system is adopted. Here, $e$, $m_e$, $h$, and $k_B$ denote the elementary charge, electron mass, Planck constant, and Boltzmann constant, respectively. In this paper, synchrotron emission from electrons in the ultra-relativistic regime is considered, and the corresponding emissivity is given by \cite{dexter2016public,pandya2016polarized,huang2024coport}
\begin{equation}
	j_{\nu} = \frac{\sqrt{3}\, e^{3} b \sin{\varphi_b}}{4 \pi m_e c^{2}} \int_{0}^{\infty} d\gamma \, \mathcal{N}(\gamma) F\left( \frac{\nu}{\nu_s} \right),\label{eq:jv}
\end{equation}
where $\gamma = 1/\sqrt{1-\beta^2}$ is the electron Lorentz factor. The function $F(x)$ is defined as
\begin{equation}
	F(x) = x \int_x^\infty dy \, K_{5/3}(y),
\end{equation}
where $K_n(y)$ is the modified Bessel function. The pitch angle $\varphi_b$ is the angle between the local magnetic-field direction $e_{(b)}^{\mu}$ and the photon propagation direction $e_{(k)}^{\mu}$
\begin{equation}
	\varphi_b = \arccos\left(e_{(b)}^{\mu} \cdot e_{(k)}^{\mu}\right),
\end{equation}
where
\begin{equation}
	e_{(k)}^{\mu} = -\left(\frac{k^{\mu}}{u^{\nu} k_{\nu}} + u^{\mu}\right), \quad 
	e_{(b)}^{\mu} = \frac{b^{\mu}}{b}.
\end{equation}
In Eq.~(\ref{eq:jv}), $\nu_s$ is the characteristic frequency,
\begin{equation}
	\nu_s = \frac{3 e b \sin\varphi_b \gamma^2}{4\pi m_ec}.
\end{equation}
For a thermal distribution, the electron distribution function is
\begin{equation}
	\mathcal{N}(\gamma) = \frac{N_{e} \gamma^{2} \beta}{\zeta_{e} K_{2}(1/\zeta_{e})} e^{(-\gamma/\zeta_{e})}.
\end{equation}
Here, $N_{e}$ is the electron number density, $\zeta_{e} = k_{B} T_{e} / m_{e} c^{2}$ is the dimensionless electron temperature, and $T_{e}$ is the thermodynamic temperature of the electrons. In the ultra-relativistic limit ($\beta\approx1,\zeta_{e} \geqslant 1$), one has $K_{2}(1/\zeta_{e}) \approx 2 \zeta_{e}^{2}$. Letting $Q = \gamma / \zeta_{e}$, one obtains
\begin{equation}
	j_{\nu} = \frac{\sqrt{3} N_{e} e^{3} b \sin{\varphi_b}}{8 \pi m_{e} c^{2}} \int_{0}^{\infty} dQ \, Q^{2} e^{-Q} \, F\left( \frac{\nu}{\nu_{s}} \right).
\end{equation}
By introducing $x = (\nu / \nu_{s}) Q^{2}$, the emissivity can be rewritten as
\begin{equation}
	j_{\nu} = \frac{N_{e} e^{2} \nu}{2 \sqrt{3} c \zeta_{e}^{2}} F(x), \quad x = \frac{\nu}{\nu_{c}}, \quad \nu_{c} = \frac{3 e b \sin{\varphi_b} \zeta_{e}^{2}}{4 \pi m_{e} c},\label{eq:em}
\end{equation}
where
\begin{equation}
	F(x) = \frac{1}{x} \int_{0}^{\infty} Q^{2} e^{-Q} F\left( \frac{x}{ Q^{2}} \right)
\end{equation}
cannot be expressed in terms of elementary functions. Its fitting formula is~\cite{mahadevan1996harmony}
\begin{equation}
	F(x) =
	2.5651 \left( 1 + 1.92\,x^{-1/3} + 0.9977\,x^{-2/3} \right)
	\exp\!\left( -1.8899\,x^{1/3} \right).
\end{equation}
The above radiation model is also referred to as anisotropic radiation, for which the magnetic field is taken to be
\begin{equation}
	b^{\mu} = \left(-\frac{u_{\phi}}{u_{t}}, 0, 0, 1\right).
\end{equation}

If the effect of the angle between the magnetic field and the emitted photon is neglected, the radiation can be treated as isotropic. For isotropic radiation, only the magnetic field strength is retained, while its direction is ignored. The corresponding emissivity can then be written as
\begin{equation}
	\hat{j}_{\nu} = \frac{1}{2} \int_{0}^{\pi} j_{\nu} \sin\varphi_b\,d \varphi_b,
	\label{eq:em2}
\end{equation}
with the fitting formula
\begin{equation}
	\hat{j}_{\nu} = \frac{N_e e^{2} \nu}{2\sqrt{3}\,c\,\zeta_{e}^{2}}\,F(x),
	\quad x = \frac{\nu}{\nu_{c}}, \quad 
	\nu_{c} = \frac{3 e b \zeta_{e}^{2}}{4\pi m_{e} c},
\end{equation}
where the fitting function for $F(x)$ is~\cite{leung2011numerical}
\begin{equation}
	F(x) =
	\frac{4.0505}{x^{1/6}}
	\left( 1 + \frac{0.4}{x^{1/4}} + \frac{0.5316}{x^{1/2}} \right)
	\exp\!\left( -1.8899\,x^{1/3} \right).
\end{equation}
Finally, for a thermal electron distribution, the absorption process follows Kirchhoff's law. The absorption coefficient can therefore be expressed as
\begin{equation}
	\alpha_{\nu} = \frac{j_{\nu}}{\hat{b}_{\nu}},
\end{equation}
where
\begin{equation}
	\hat{b}_{\nu}=\frac{2h\nu^{3}}{c^{2}}\left[\exp\left(h\nu/k_{B}T_{e}\right)-1\right]^{-1}
\end{equation}
is the Planck blackbody function.

\begin{figure}[!htbp]
	\centering
	\subfigure[$N=1.2,\theta_o=0^\circ$]{\includegraphics[width=0.28\textwidth]{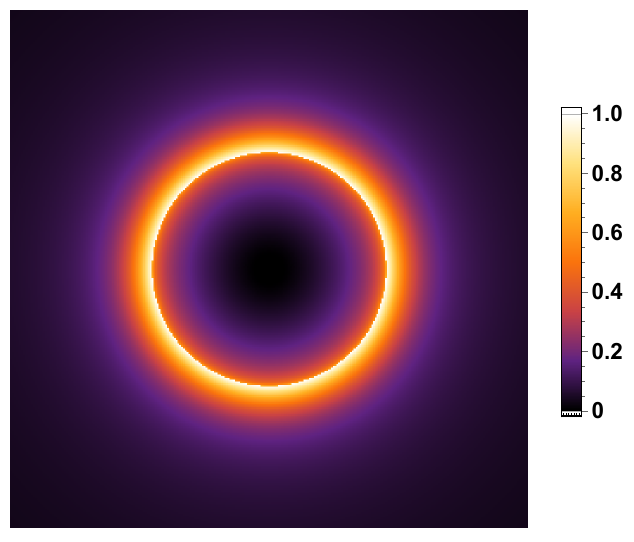}}
	\subfigure[$N=1.2,\theta_o=17^\circ$]{\includegraphics[width=0.28\textwidth]{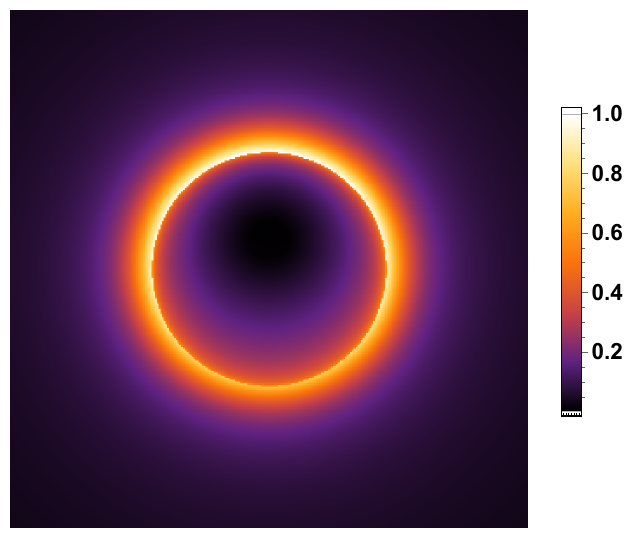}}
	\subfigure[$N=1.2,\theta_o=80^\circ$]{\includegraphics[width=0.28\textwidth]{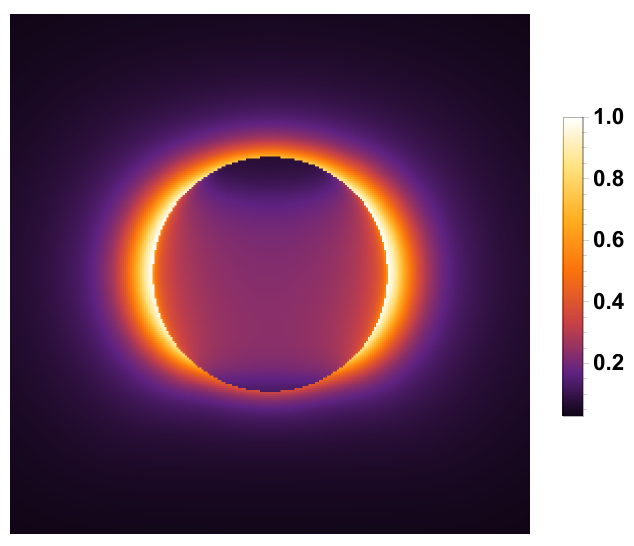}}
	
	\subfigure[$N=1.3,\theta_o=0^\circ$]{\includegraphics[width=0.28\textwidth]{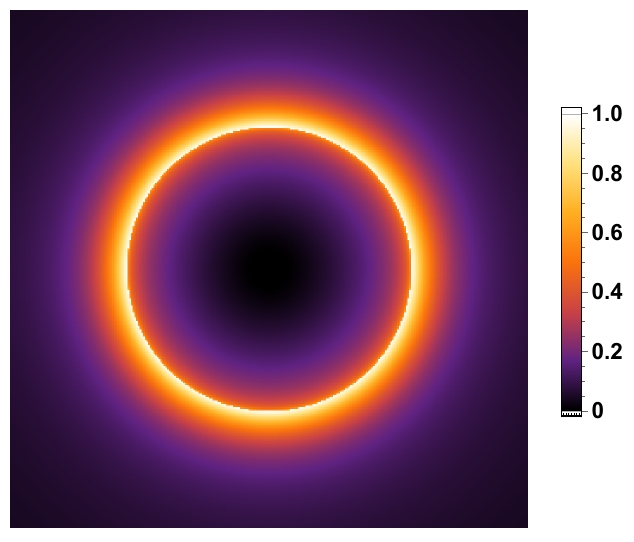}}
	\subfigure[$N=1.3,\theta_o=17^\circ$]{\includegraphics[width=0.28\textwidth]{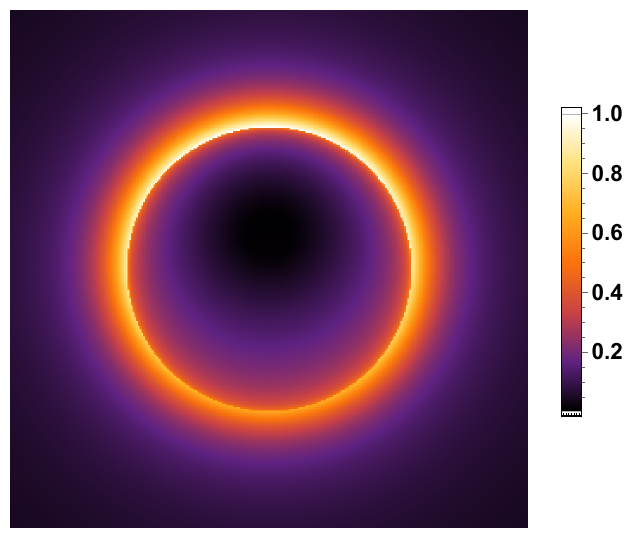}}
	\subfigure[$N=1.3,\theta_o=80^\circ$]{\includegraphics[width=0.28\textwidth]{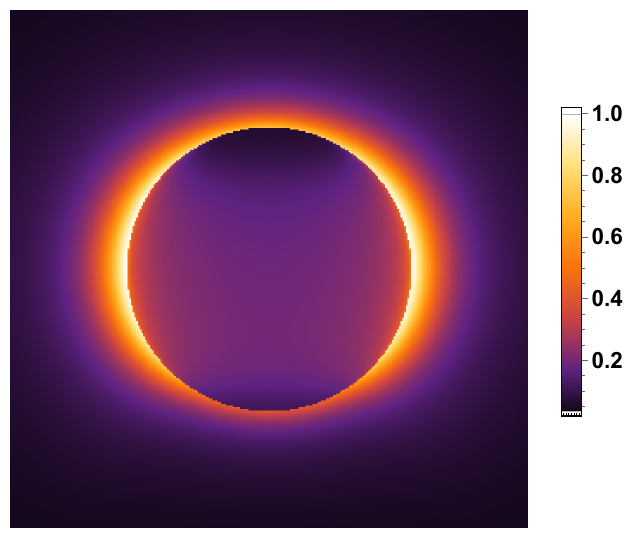}}
	
	\subfigure[$N=1.4,\theta_o=0^\circ$]{\includegraphics[width=0.28\textwidth]{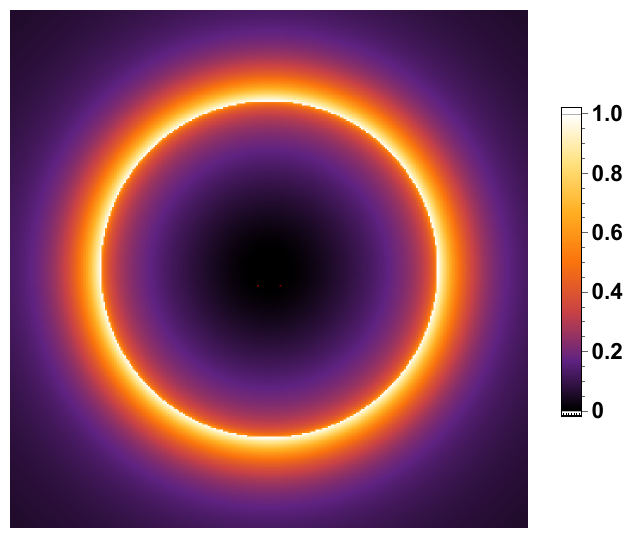}}
	\subfigure[$N=1.4,\theta_o=17^\circ$]{\includegraphics[width=0.28\textwidth]{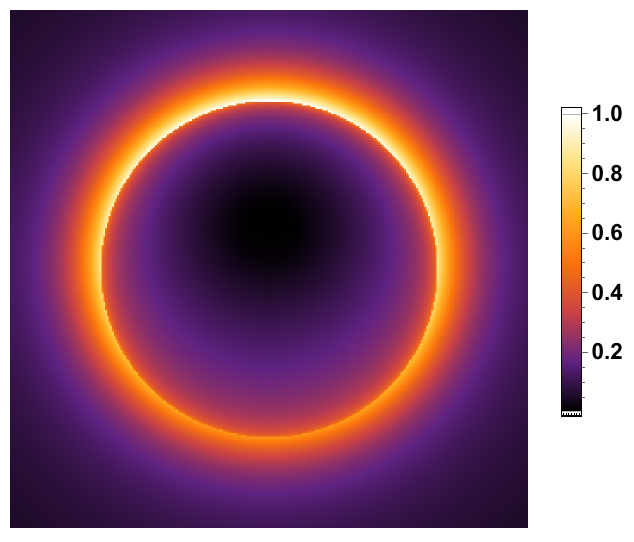}}
	\subfigure[$N=1.4,\theta_o=80^\circ$]{\includegraphics[width=0.28\textwidth]{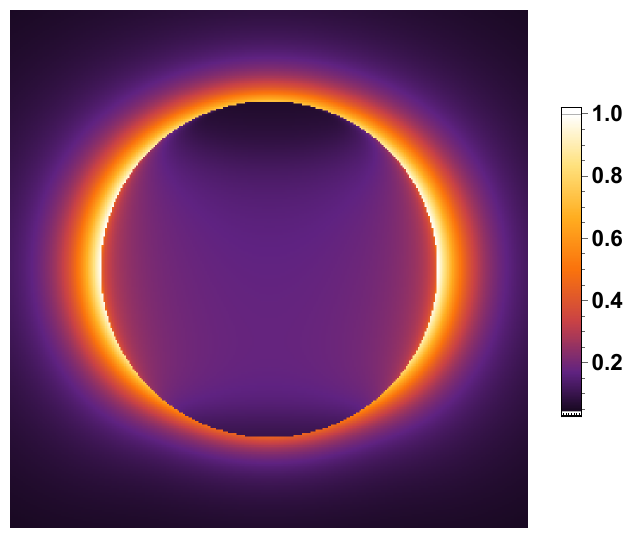}}
	
	\caption{Effects of the polytropic index $N$ and the observer inclination angle $\theta_o$ on the optical images of neutron stars in the case of isotropic radiation.}
	\label{fig1}
\end{figure}

\section{Numerical Results}\label{sec4}

This section presents the numerical results. Fig.~\ref{fig1} illustrates the effects of the polytropic index $N$ and the observer inclination angle $\theta_o$ on the optical images of neutron stars in the case of isotropic radiation. In all images, a bright closed ring-like structure can be identified, which is referred to as the higher order image. The higher order image is formed by photons that orbit the star one or more times before reaching the observer, and it arises from strong gravitational lensing. Similar to the black hole case, the strong gravitational field of a neutron star can deflect part of the photon trajectories and allow them to reach the observer after passing around the star. Outside the higher order image, the region where the intensity remains nonzero is referred to as the primary image, which corresponds to photons propagating directly from the accretion disk to the observer. It is worth noting that there exists a region of significantly reduced intensity inside the higher order image. This region is related to the truncation condition adopted in this paper and reflects the obscuration of electromagnetic radiation by the neutron star interior. For a geometrically thin accretion disk, the accreting material exists only on the equatorial plane and extends inward to the stellar radius, so the neutron star silhouette appears in the image as a black region with a well-defined boundary~\cite{yang2026distinguishing}. For a geometrically thick accretion disk, however, this region may be partially obscured by radiation originating outside the equatorial plane.

In Fig.~\ref{fig1}, when $\theta_o = 0^\circ$, the higher order image appears as a perfect ring. As $\theta_o$ increases to $17^\circ$, the low intensity region inside the higher order image shifts upward on the screen. When $\theta_o$ is further increased to $80^\circ$, the overall intensity inside the higher order image becomes enhanced, and only relatively pronounced dark regions remain in the upper and lower parts. This behavior reflects the obscuration of the neutron star silhouette by accreting material located away from the equatorial plane. On the other hand, a comparison among different rows shows that increasing the polytropic index $N$ significantly enlarges the size of the higher order image, while having almost no effect on its shape.

\begin{figure}[!htbp]
	\centering 
	\subfigure[$N=1.2,\theta_o=0^\circ$]{\includegraphics[width=0.28\textwidth]{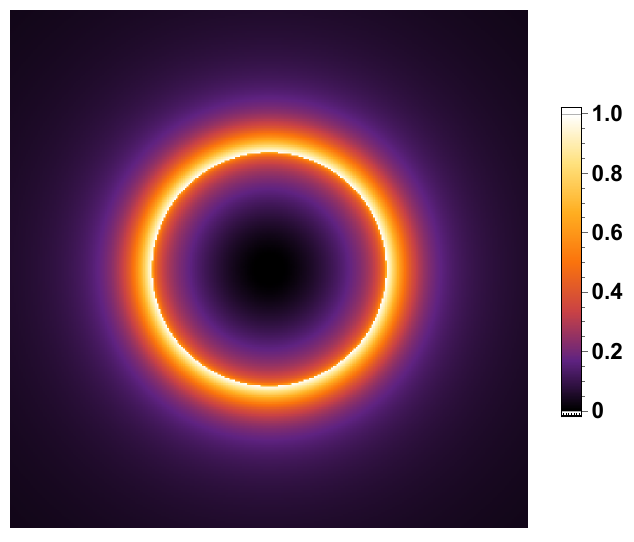}}
	\subfigure[$N=1.2,\theta_o=17^\circ$]{\includegraphics[width=0.28\textwidth]{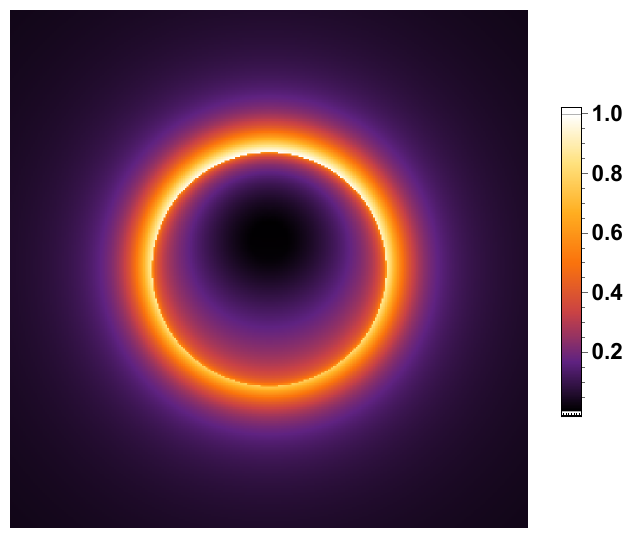}}
	\subfigure[$N=1.2,\theta_o=80^\circ$]{\includegraphics[width=0.28\textwidth]{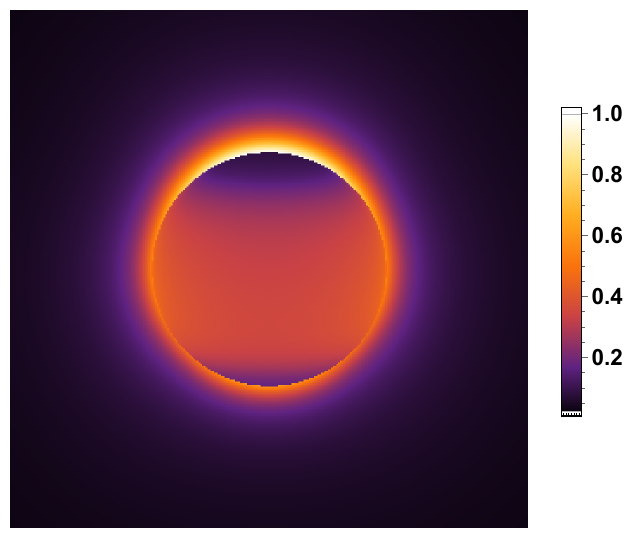}}
	
	\subfigure[$N=1.3,\theta_o=0^\circ$]{\includegraphics[width=0.28\textwidth]{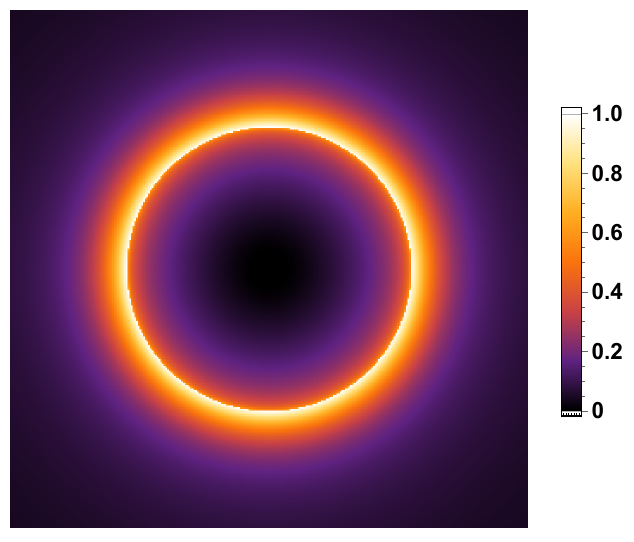}}
	\subfigure[$N=1.3,\theta_o=17^\circ$]{\includegraphics[width=0.28\textwidth]{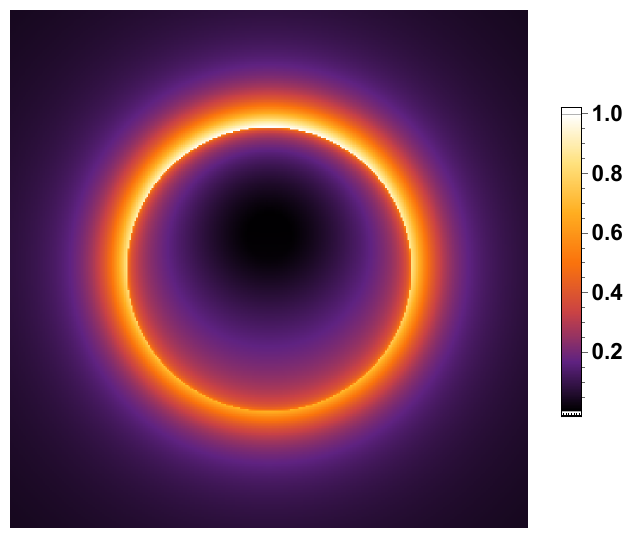}}
	\subfigure[$N=1.3,\theta_o=80^\circ$]{\includegraphics[width=0.28\textwidth]{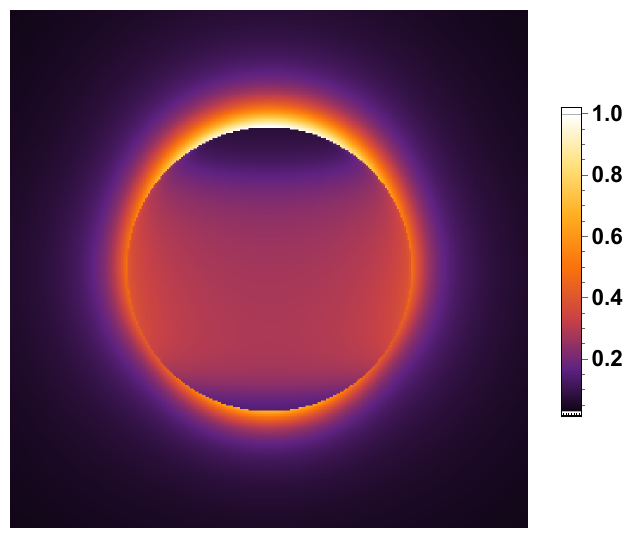}}
	
	\subfigure[$N=1.4,\theta_o=0^\circ$]{\includegraphics[width=0.28\textwidth]{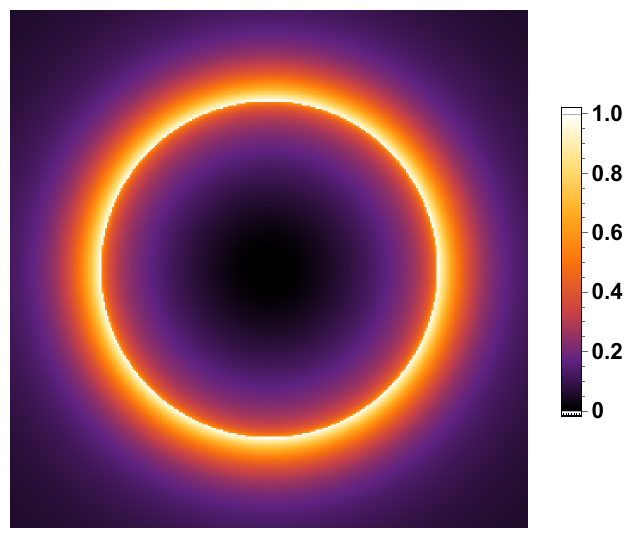}}
	\subfigure[$N=1.4,\theta_o=17^\circ$]{\includegraphics[width=0.28\textwidth]{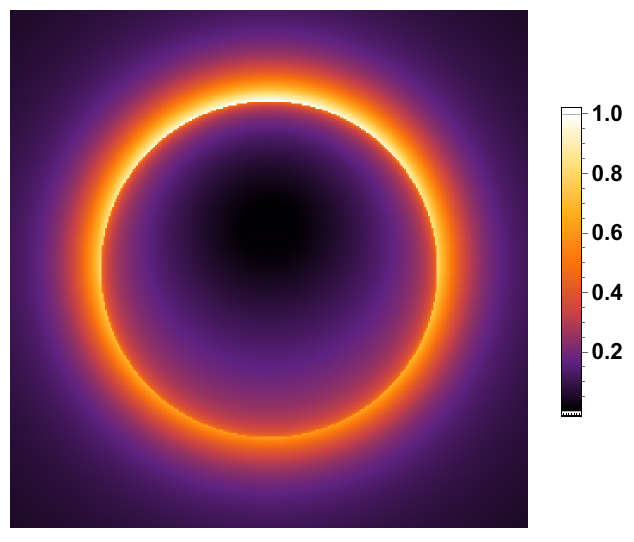}}
	\subfigure[$N=1.4,\theta_o=80^\circ$]{\includegraphics[width=0.28\textwidth]{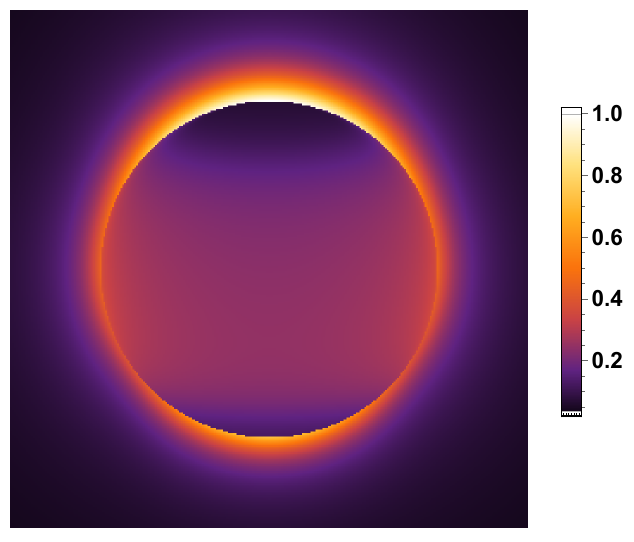}}
	
	\caption{Effects of the polytropic index $N$ and the observer inclination angle $\theta_o$ on the optical images of neutron stars in the case of anisotropic radiation.}
	\label{fig2}
\end{figure}

Fig.~\ref{fig2} shows the effects of $N$ and $\theta_o$ on the intensity distribution in the case of anisotropic radiation. For small observer inclination angles, the resulting optical images are essentially the same as those in the isotropic radiation case. However, for large observer inclination angles, the dark region inside the higher order image nearly disappears and remains only faintly visible in the upper and lower parts. This indicates that, under the combined effect of anisotropic radiation and a large observer inclination angle, the contribution from radiation away from the equatorial plane is significantly enhanced. Meanwhile, increasing $N$ still leads to an enlargement of the higher order image.

\begin{figure}[!htbp]
	\centering 
	\subfigure[NS,Isotropic]{\includegraphics[width=0.23\textwidth]{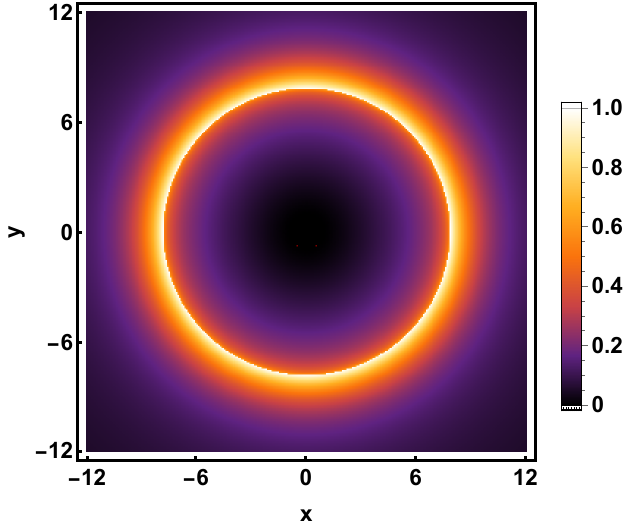}}
	\subfigure[NS,Anisotropic]{\includegraphics[width=0.23\textwidth]{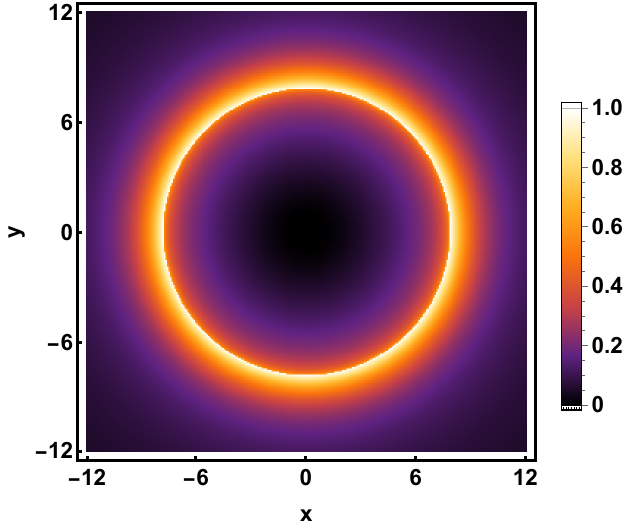}}
	\subfigure[BH,Isotropic]{\includegraphics[width=0.23\textwidth]{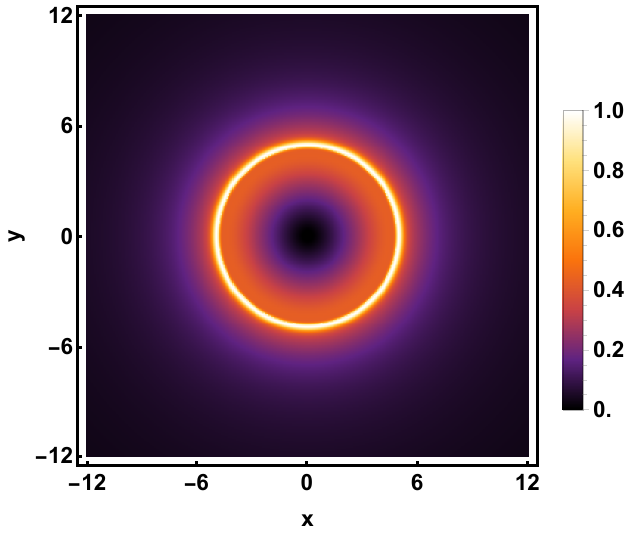}}
	\subfigure[BH,Anisotropic]{\includegraphics[width=0.23\textwidth]{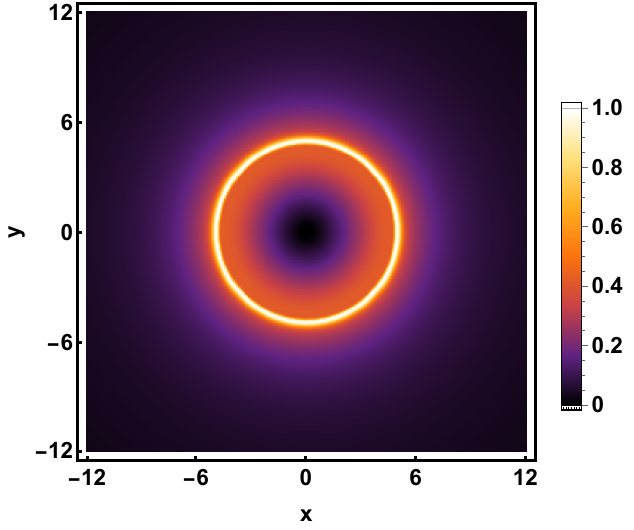}}

	\subfigure[NS,Isotropic]{\includegraphics[width=0.23\textwidth]{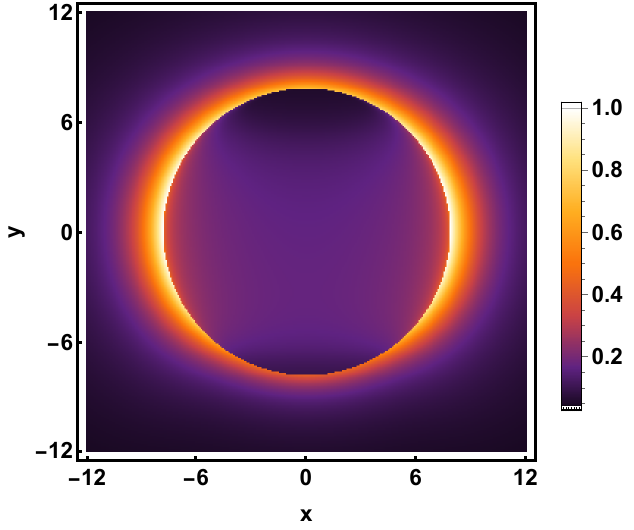}}
	\subfigure[NS,Anisotropic]{\includegraphics[width=0.23\textwidth]{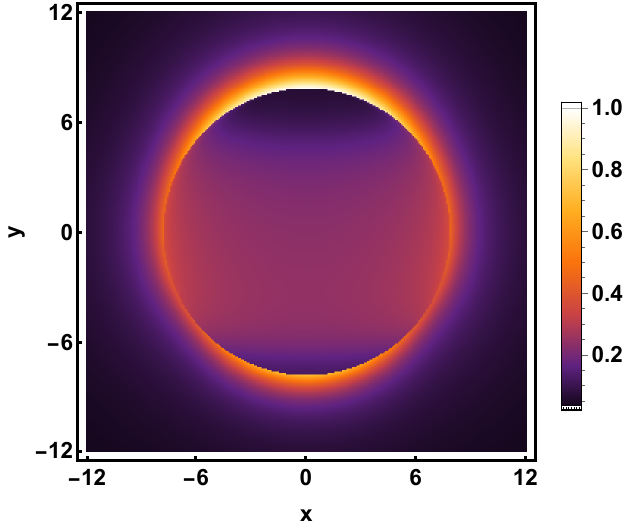}}
	\subfigure[BH,Isotropic]{\includegraphics[width=0.23\textwidth]{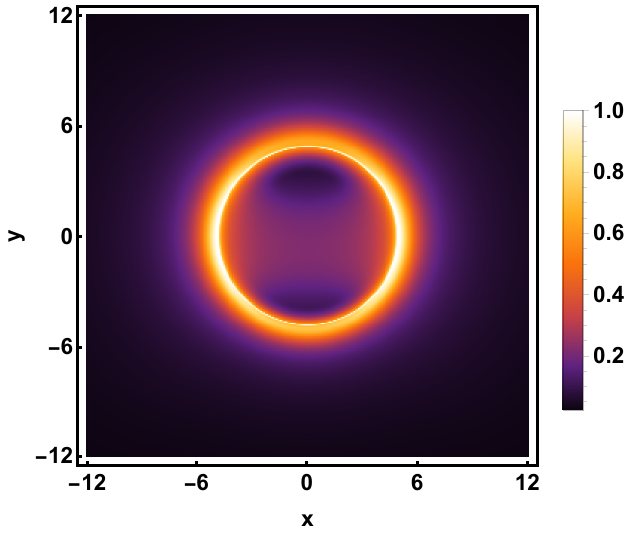}}
	\subfigure[BH,Anisotropic]{\includegraphics[width=0.23\textwidth]{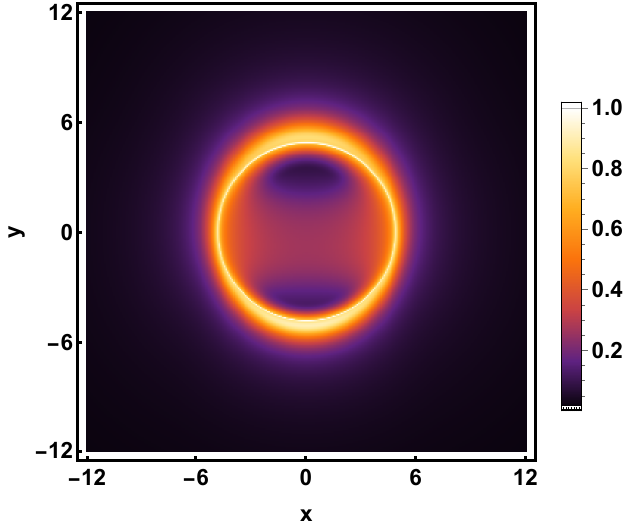}}
	
	\caption{Comparison of the optical images of a neutron star (NS) and a Schwarzschild black hole (BH) in the RIAF model for different radiation mechanisms. The first and second rows correspond to observer inclination angles of $\theta_o=0^\circ$ and $80^\circ$, respectively. The polytropic index of the neutron star is fixed at $N=1.4$, while all other plotting parameters are kept the same.}
	\label{fig3}
\end{figure}

To compare the differences between the optical images of neutron stars and Schwarzschild black holes, Fig.~\ref{fig3} presents the corresponding images of both objects in the RIAF model. Although the exterior spacetime of the neutron star is also described by the Schwarzschild metric, clear differences are still visible between the two sets of images. First, in all cases, a bright ring corresponding to the higher order image can be identified. However, because the Schwarzschild black hole is more compact than the neutron star, the radius of its higher order image is significantly smaller. At lower observer inclination angles (the first row), a pronounced low intensity region appears inside the higher order image in both cases. For the neutron star, this region occupies a larger fraction of the interior of the higher order image than in the black hole case, and the radiation type has only a minor effect on the image structure in this regime. It is worth noting that the dark region inside the black hole image originates from the event horizon, whereas the corresponding dark region in the neutron star image arises from the truncation assumption. At higher observer inclination angles (the second row), the higher order image of the black hole is more clearly distinguishable than that of the neutron star. In this case, the influence of radiation originating outside the equatorial plane is stronger in the anisotropic radiation case than in the isotropic radiation case.

\section{Conclusions}\label{sec5}
Compared with thin disks, thick disks are better suited to capturing realistic astrophysical environments. In this paper, we have therefore investigated the optical images of neutron stars with a polytropic equation of state in the presence of a geometrically thick and optically thin accretion disk, and compared them with the shadow of a Schwarzschild black hole. First, a smooth fitted metric was constructed for both the interior and exterior regions of the neutron star, thereby laying the foundation for the imaging calculations in combination with the ray-tracing method. The RIAF model and two electron radiation mechanisms, namely isotropic radiation and anisotropic radiation, were then introduced. By numerically solving the geodesic equations and the radiative transfer equation, the corresponding optical images of neutron stars were obtained.

Within the above imaging framework, we assume that photon trajectories are terminated once they reach the neutron star surface. Although neutron stars do not possess an event horizon and photons can in principle propagate into the stellar interior, the present work focuses primarily on gravitational effects and does not take into account physical processes such as refraction and transmission inside the star. This truncation treatment leads to a slightly lower brightness than in the case where photons are fully allowed to propagate into the stellar interior. Nevertheless, the overall image morphology and its key structures still differ significantly from those of a black hole shadow, which is sufficient to demonstrate the essential difference in the imaging mechanisms of neutron stars and black holes.

The numerical results exhibit an overall structure similar to that seen in the EHT observational images. All models show an outer bright ring and an inner region of reduced intensity, with the outer bright ring corresponding to the higher order image. An increase in the polytropic index $N$ enlarges the size of the higher order image, while having only a limited effect on its shape. As the observer inclination angle $\theta_o$ increases, the obscuration of the neutron star silhouette by radiation outside the equatorial plane becomes progressively stronger, and this effect is more pronounced in the anisotropic radiation case. Compared with the black hole shadow, under the same parameter configuration, the neutron star has a larger higher order image and a more extended inner region of reduced intensity, whereas the higher order image of the black hole is more readily distinguishable. At large observer inclination angles, the inner low intensity regions of both objects become obscured and split into upper and lower parts.

The study of neutron star optical images in the presence of geometrically thick accretion disks provides a new basis for distinguishing neutron stars from black holes in high-resolution astronomical observations. This work adopts a polytropic equation of state to simplify numerical solutions and imaging calculations, but the framework is based on numerical solutions of the stellar structure and the matching of the exterior metric, and can be straightforwardly extended to more realistic neutron star models. By incorporating the corresponding equations of state of these models into the current framework and generating the required interior structure data via interpolation or fitting, one can perform ray-tracing and radiative transfer calculations to produce the resulting images.

For the model adopted in this work, if there existed an alternative black hole beyond Schwarzschild whose photon ring and central dark region appeared similar to those of a neutron star image, it would be difficult to distinguish the two based on optical images alone. Although the underlying physics are different, the scale of a black hole shadow is primarily determined by the spacetime structure, whereas the central dark region of a neutron star is also affected by the stellar radius and the surface boundary condition. These differences, however, are not directly reflected in optical images. This is mainly because the higher order images of both neutron stars and black holes are governed by strong gravitational lensing. Radiation from outside the disk plane reduces the brightness contrast between the bright ring and the dark region, leading to an observational degeneracy. Nevertheless, the two objects may still be distinguished through polarimetric imaging. The polarization pattern of a black hole photon ring exhibits a self-similar structure determined by general relativity~\cite{himwich2020universal}. In contrast, for neutron star surface emission, vacuum polarization in strong magnetic fields and rotation of the Stokes parameters can produce pronounced asymmetric variations in the observed polarization angle with rotational phase~\cite{taverna2015polarization}. A more fundamental distinction is that a neutron star has a physical surface, so further discrimination should focus on the effects of the surface on photon propagation. In this work, rays reaching the neutron star surface are terminated, which is equivalent to adopting a fully absorbing surface boundary condition. In reality, however, the surface may involve more complex behavior, such as delayed re-emission, reflection, and transmission~\cite{garcia2022relativistic,carballo2023constraints}. Including these effects can help distinguish the two types of images. In addition, hotspots may exist on the neutron star surface. They can arise, for example, from X-ray bursts or from locally heated regions near the magnetic poles. The localized bright regions produced by such hotspots in the image provide observable signatures for identifying neutron stars~\cite{goodwin2021x}. Finally, we note that although the RIAF model captures the main radiative properties of the accretion flow, it does not include the full GRMHD processes. Incorporating these processes into simulations can play an important role in image discrimination. In future work, we plan to introduce full GRMHD simulations and systematically evaluate the effects of thick disk radiation, polarization, and complex surface behavior on optical imaging. Through these studies, we aim to provide a robust theoretical basis for distinguishing neutron stars from black holes in high-resolution observations.


\appendix 
\section{Comparison of Accretion Flow Motion Modes}\label{appendix1}

In the main text, the accretion flow was restricted to purely infalling motion. Here we briefly introduce the orbiting motion and combined motion for comparison.

\textbf{Orbiting motion} \cite{gold2020verification}

In the orbiting motion, the fluid rotates around the neutron star, and the four-velocity has only $u^t$ and $u^\phi$ components
\begin{equation}
	u^\mu = u^t (1, 0, 0, \Omega),
\end{equation}
where
\begin{equation}
	u^t = \sqrt{-\frac{1}{g_{tt} + g_{\phi\phi} \Omega^2}}, \quad
	\Omega = -\frac{g_{tt} l}{g_{\phi\phi}}, \quad
	l = -\frac{u_\phi}{u_t} = \frac{R^{3/2}}{1 + R}, \quad
	R = r \sin\theta.
\end{equation}
Here, $l$ denotes the specific angular momentum of the fluid.

\textbf{Combined motion} \cite{pu2016effects}

The combined motion represents a weighted superposition of infalling and orbiting components. For the infalling component,
\begin{equation}
	\Omega_f = 0, \quad u^r_f = -\sqrt{-(1 + g^{tt}) g^{rr}}.
\end{equation}
For the orbiting component,
\begin{equation}
	R = r \sin\theta, \quad l = \frac{R^{3/2}}{1 + R}, \quad \Omega_o = -\frac{g_{tt} l}{g_{\phi\phi}}.
\end{equation}
The combination is expressed as
\begin{equation}
	\Omega = \Omega_o + \beta_1 (\Omega_f - \Omega_o), \quad
	u^r = \beta_2 u^r_f, \quad
	u^t = \sqrt{-\frac{1 + g_{rr} (u^r)^2}{g_{tt} + g_{\phi\phi} \Omega^2}}.
\end{equation}
Thus, the four-velocity of the combined motion is
\begin{equation}
	u^\mu = (u^t, u^r, 0, u^t \Omega).
\end{equation}
The parameters $\beta_1$ and $\beta_2$ control the weighting of the orbiting and infalling components. In this work, we set $\beta_1 = \beta_2 = 0.5$.

\begin{figure}[!htbp]
	\centering
	\subfigure[Orbiting motion]{\includegraphics[width=0.28\textwidth]{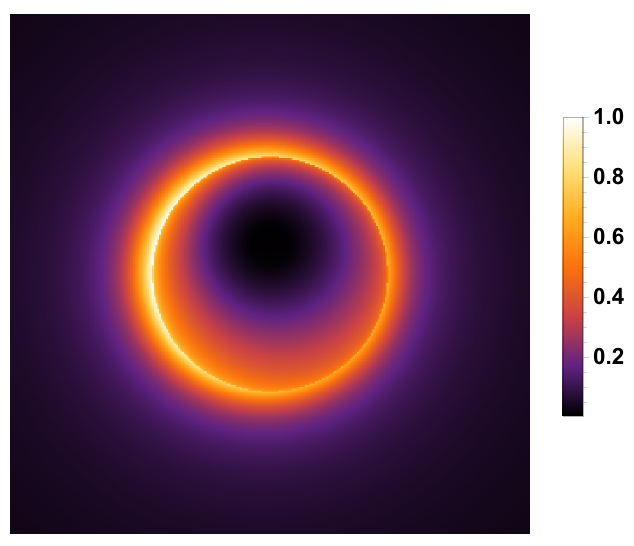}}
	\subfigure[Infalling motion]{\includegraphics[width=0.28\textwidth]{Iso2}}
	\subfigure[Combined motion]{\includegraphics[width=0.28\textwidth]{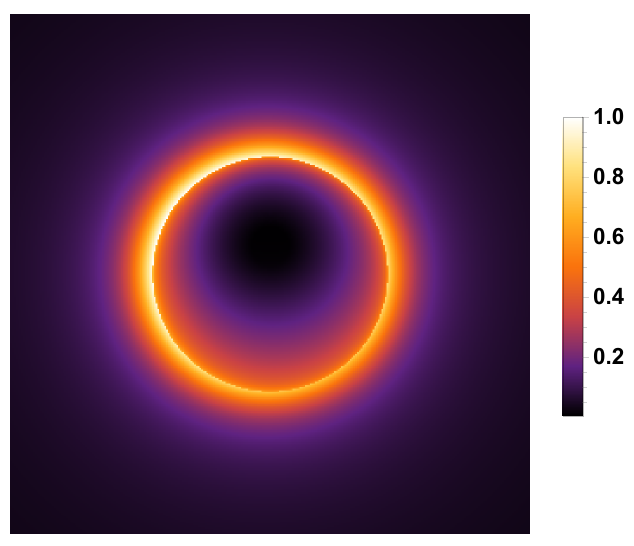}}
	
	
	\caption{Effect of different accretion flow motion modes on the optical images under isotropic radiation. The fixed parameters are $N=1.2$ and $\theta_o = 17^\circ$.}
	\label{fig4}
\end{figure}

Fig.~\ref{fig4} illustrates the effect of different accretion flow motion modes on the optical images of a neutron star. In all three cases, the images exhibit a central dark region surrounded by a bright ring. Compared with purely infalling motion, the presence of an orbiting component increases both the obscuration of the neutron star silhouette by material off the equatorial plane and the asymmetry of the optical images, which results in a smaller central dark region.

To focus on the main morphological features highlighted above, all optical images presented in the main text are computed using purely infalling motion. This choice is motivated by two considerations. First, from a physical perspective, in the RIAF model the radial infall of the accreting material minimizes the obscuration from material off the equatorial plane, resulting in images that are closer to a spherically symmetric configuration in projection. Second, from a computational perspective, neglecting the orbital component ensures stability in the ray-tracing and radiative transfer calculations.

\noindent {\bf Acknowledgments}

\noindent This work is supported by the National Natural Science Foundation of China (Grants Nos. 12375043, 12575069) and the Chongqing Normal University Fund Project (Grant No. 26XLB001).

\bibliographystyle{utphys} 
\bibliography{biblio} 

\end{document}